\documentclass[12pt]{iopart}

\usepackage{iopams}
\usepackage{graphicx}
\begin{document}

\title{Direct photons at low $p_T$ measured in PHENIX}

\author{D.~Peressounko for the PHENIX collaboration\footnote[7]{For the full list of PHENIX authors and acknowledgements, see Appendix
'Collaborations' of this volume}}

\address{RRC "Kurchatov Institute", Kurchatov sq.1, Moscow, 123182, Russia}

\begin{abstract}
Direct photon spectra measured at small $p_T$ in p+p, d+Au and Au+Au
collisions at $\sqrt{s_{NN}}=200$ GeV are presented. Several
measurement techniques including statistical subtraction, tagging,
and internal and external conversion were applied and found to
produce consistent results. The p+p and d+Au results are found to be
in very good agreement with pQCD predictions over the entire $p_T$
range. No excess of direct photons in Au+Au collisions with respect
to binary scaled d+Au data is observed within systematic errors.
\end{abstract}


\section{Introduction}
Due to their extremely large mean free path length direct photons are
considered as an excellent tool to explore the initial state of p+p
and A+A collisions, to study nucleon structure functions and their
modification in nuclei. But the most exciting application of direct
photons is the possibility to deduce the temperature of the hot
matter created in A+A collisions and to extract its equation of
state. Due to the power-law form of the spectrum of hard prompt
photons they will dominate at high $p_T$ while matter-related direct
photons can be observed only in the soft $p_T$ region. Despite the
considerable progress in the theoretical description of direct photon
production in p+p collisions in the last 10 years \cite{Vogelsang},
theoretical calculations are still not very reliable at small photon
momenta and to account for the prompt photon contribution in A+A
collisions we need a baseline -- direct photon spectra measured in
p+p and d+A collisions.

Direct photon extraction is an extremely complicated experimental
task, especially at small $p_T$, because the contribution of direct
photons is only a few percent of the total photon yield and it
rapidly decreases at small $p_T$.  The dominating photon source is
final-state hadron decays, $\pi^0\to 2\gamma$, $\eta\to 2\gamma$ etc.
Having a variety of detector subsystems, PHENIX has developed a set
of approaches with the aim of increasing the signal/background ratio
and decrease the systematic error.

\section{Data analysis and results}

\begin{figure}
\begin{center}
  \includegraphics[width=75mm]{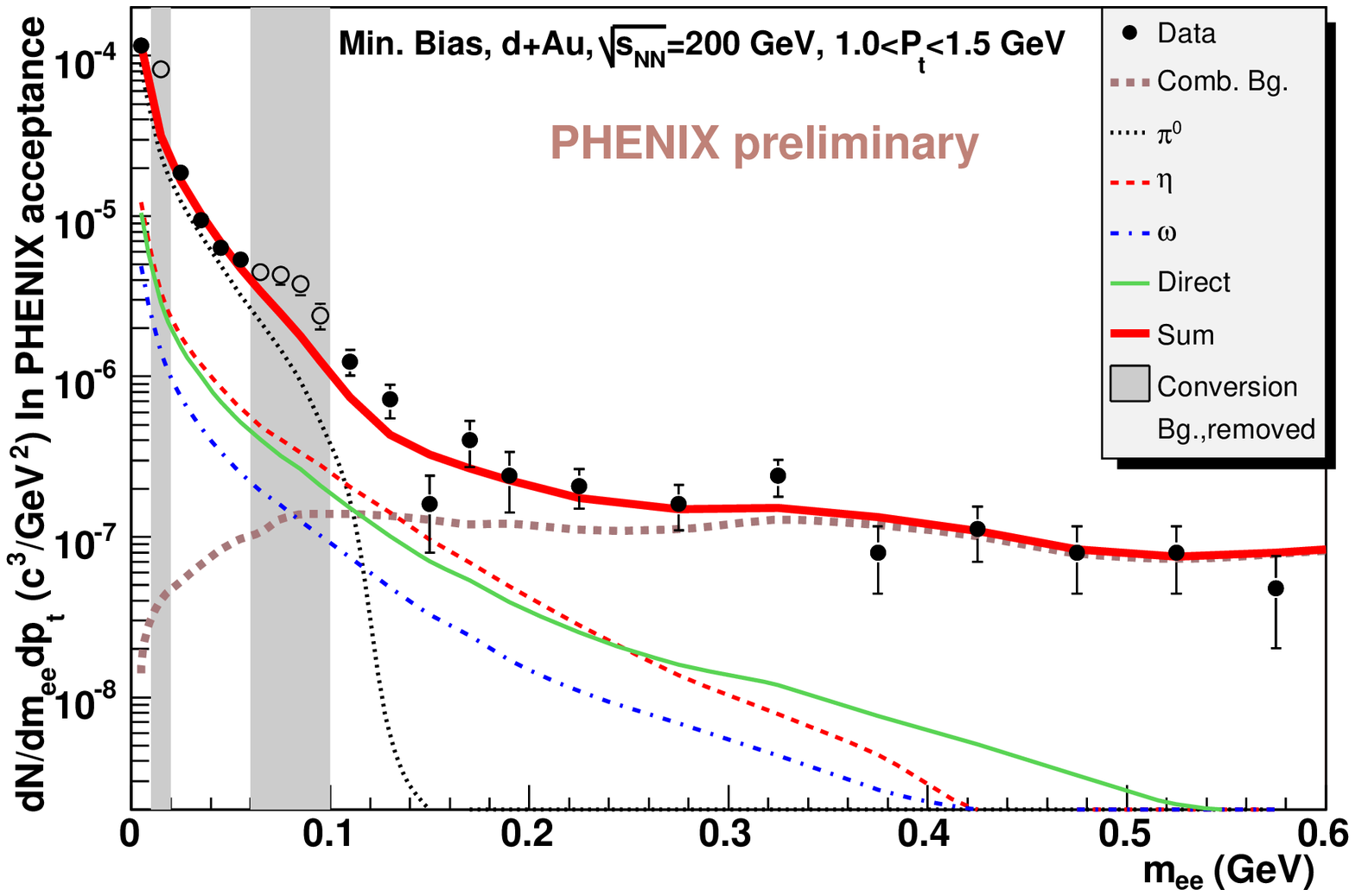}\hfill
  \includegraphics[width=75mm]{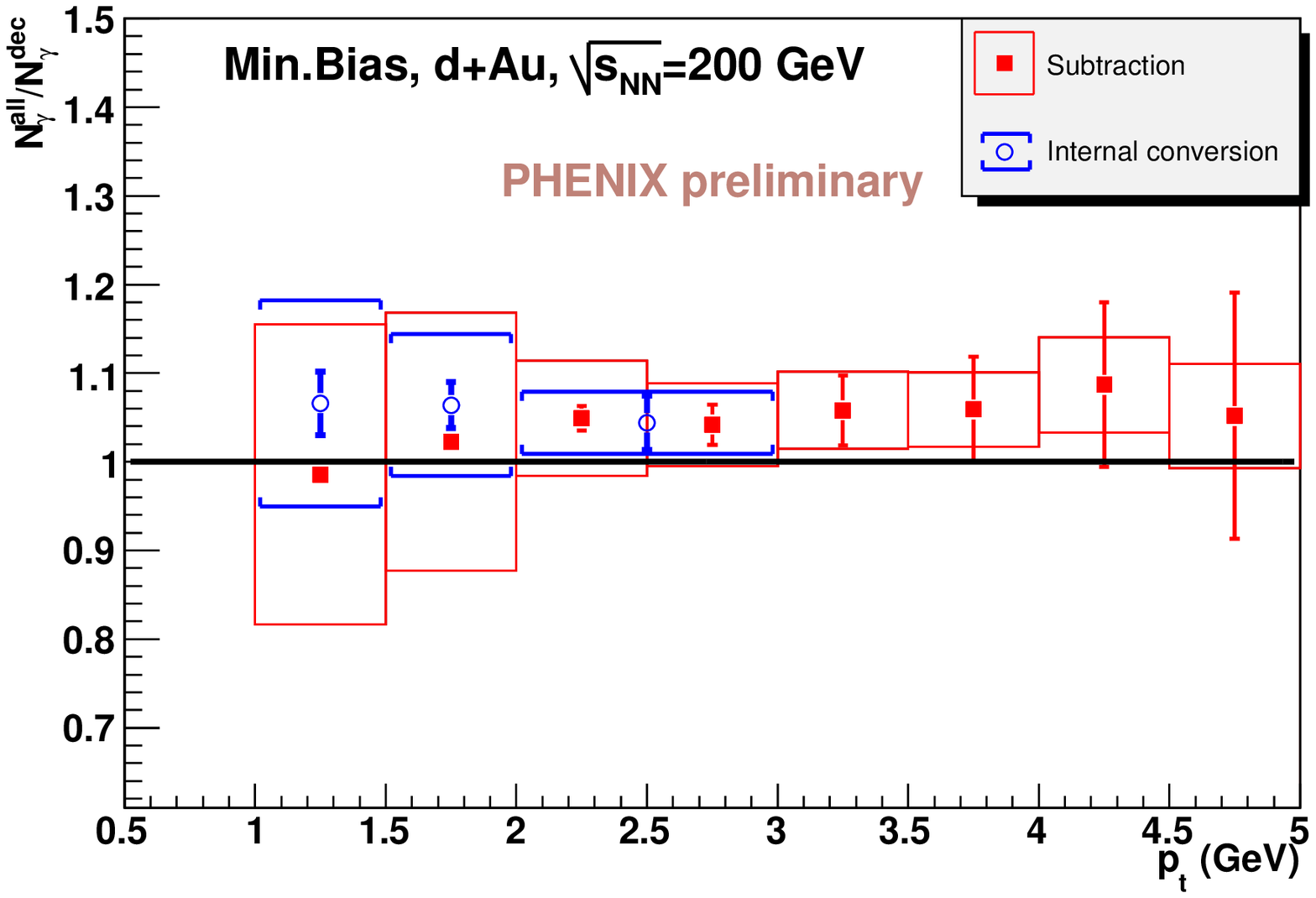}\\
  \caption{Left plot: Di-electron mass distribution, measured in minimum bias d+Au
  collisions at $\sqrt{s_{NN}}=200$ GeV in the PHENIX acceptance, decomposed
  into contributions from combinatorial background, meson Dalitz decays,
  and direct photon internal conversion. The two grey bands show regions
  contaminated by external conversion and excluded from the fit. Right plot:
  Ratio of inclusive to decay photon yields measured in same collisions by two methods.}\label{fig:mee}
\end{center}
\end{figure}

The results presented here were taken in the 2003 RHIC run 3 (p+p and
d+Au) and 2004 run 4 (Au+Au) with the PHENIX setup for each run
described in \cite{PHENIXpp} and \cite{Akiba},  respectively. The
analysis is based on integrated luminosity of 266 nb$^{-1}$ of p+p
events, $\approx 3$ billion d+Au events, and $\approx 900$ M Au+Au
events.

To produce the direct photon spectrum we measure the inclusive photon
yield and subtract the decay photon contribution.  The inclusive
photon spectrum is measured with the electromagnetic calorimeter.
Charged hadron contamination is removed by a cut on the distance to a
projection of charged hit measured in the pad chamber, while neutral
hadron contamination is estimated by analyzing the calorimeter
response to identified charged hadrons and full GEANT simulations. In
addition, very efficient shower shape and time-of-flight cuts are
applied reducing the contamination of neutral hadrons from $\sim
30$\% to $3-5$\% at low $p_T$. Finally, acceptance, efficiency,
conversion loss, and other corrections are applied to produce the
inclusive photon spectrum.

\begin{figure}
\begin{center}
  \includegraphics[width=75mm,height=7.3cm]{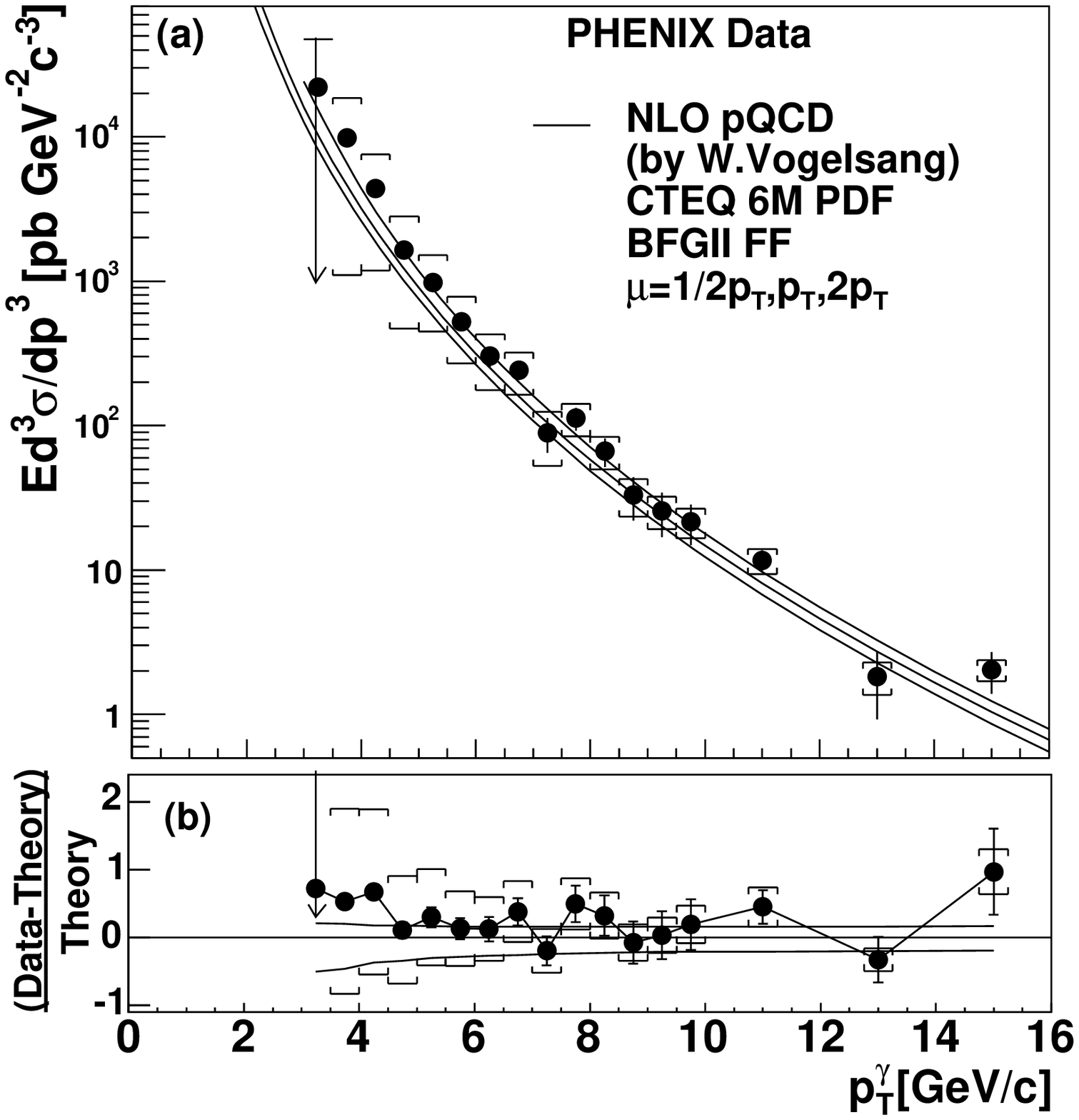}\hfill
  \includegraphics[width=75mm,height=7.2cm]{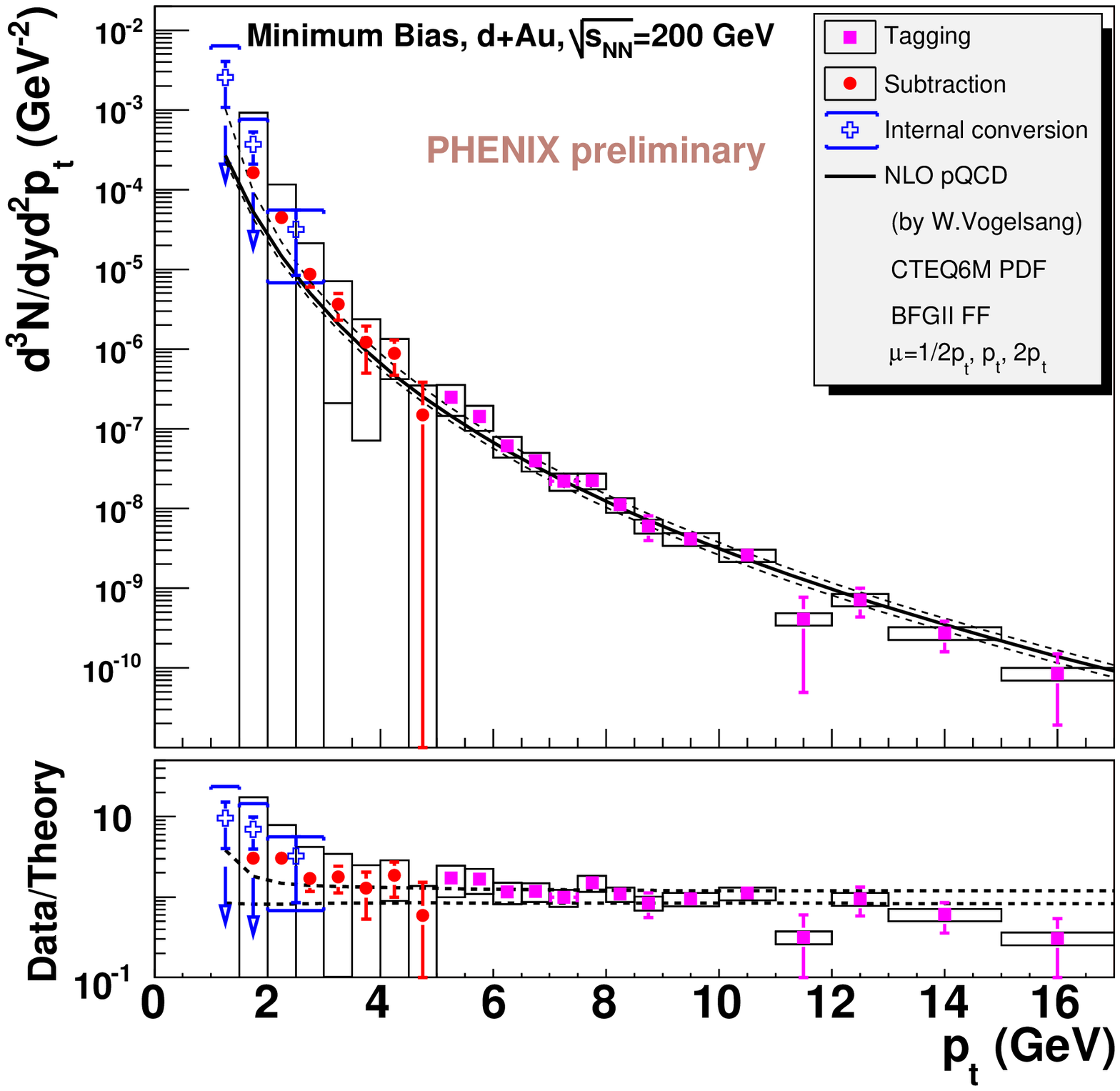}\\
  \caption{Left plot: Direct photon production cross-section measured in p+p collisions at
  $\sqrt{s_{NN}}=200$ GeV compared to pQCD predictions. Right plot: Direct
  photon yield measured by three methods in minimum bias d+Au collisions at
  $\sqrt{s_{NN}}=200$ GeV compared to binary scaled pQCD predictions
  \cite{Vogelsang}.}\label{fig:ratio}
\end{center}
\end{figure}

With the variety of detector subsystems present in PHENIX several
alternative techniques can be applied to remove decay photons. The
basic technique -- subtraction -- includes measurement of the
$\pi^0$, $\eta$ and $\omega$ meson spectra, calculation of the
spectrum of decay photons with Monte Carlo simulations, and
subtraction of the decay photon spectrum from the inclusive photon
spectrum \cite{PHENIXpp}. In another technique
 -- tagging -- photon pairs with mass close to the $\pi^0$ mass
are removed from the inclusive photon spectrum. Corrections are then
applied for  fake partners, missing partners, and heavier meson
decays contributions  \cite{PHENIXpp}. An important extension of the
tagging technique is the external conversion method. Its key feature
is the measurement of the tagged photons using external photon
conversion on the material of beam pipe.  Imposing a cut on
orientation of the $e^+e^-$ pair in the magnetic field, we remove
combinatorial background and contributions from Dalitz decays
producing a very clean inclusive photon sample.  Then decay photons
are removed applying a standard tagging technique taking partner
photons registered in the calorimeter.

Due to the small signal/background ratio both the tagging and
subtraction approaches produce relatively large systematic errors.
Errors can be considerably reduced using the internal conversion
method.  This method is based on the fact that any system emitting
real photons emits virtual photons that convert to $e^+e^-$ pairs
with a universal $m_{ee}$ distribution as $m_{ee}\to 0$
\cite{Kroll-Wada}. On the other hand, the di-electron mass cannot
exceed the mass of the decaying system. Therefore measurement of
the proportion of direct virtual photons at $m_{ee}>100$~MeV
excludes the $\pi^0\to\gamma e^+e^-$ contribution and thus
increases the signal/background ratio by $\sim 10$ times. This
technique is illustrated in Figure~\ref{fig:mee}. The di-electron
mass distribution is fitted with the sum of the measured
combinatorial background and the simulated contributions of
virtual decay and direct photons. This fit has two free
parameters: the proportion of direct photons and the absolute
normalization of the simulated contributions. It's seen that at
$m_{ee}\lesssim 100$ MeV the direct photon contribution is quite
small while at higher mass it becomes comparable with the other
contributions.

The ratio of inclusive photons over decay photons in d+Au
collisions was extracted using the internal conversion and
subtraction techniques (see Figure~\ref{fig:mee}). Good agreement
is found between the two approaches. The systematic errors in the
case of the internal conversion method are not as small as
recently reported in an analysis of Run~4 Au+Au collisions
\cite{Akiba}. The reason is that the PHENIX setup in Run~3 was not
optimized for this measurement: a huge background (dashed area in
left plot of Figure~\ref{fig:mee}) was produced due to external
conversions on the MVD vertex detector, removed in later runs.

\begin{figure}
\begin{center}
  \includegraphics[width=75mm]{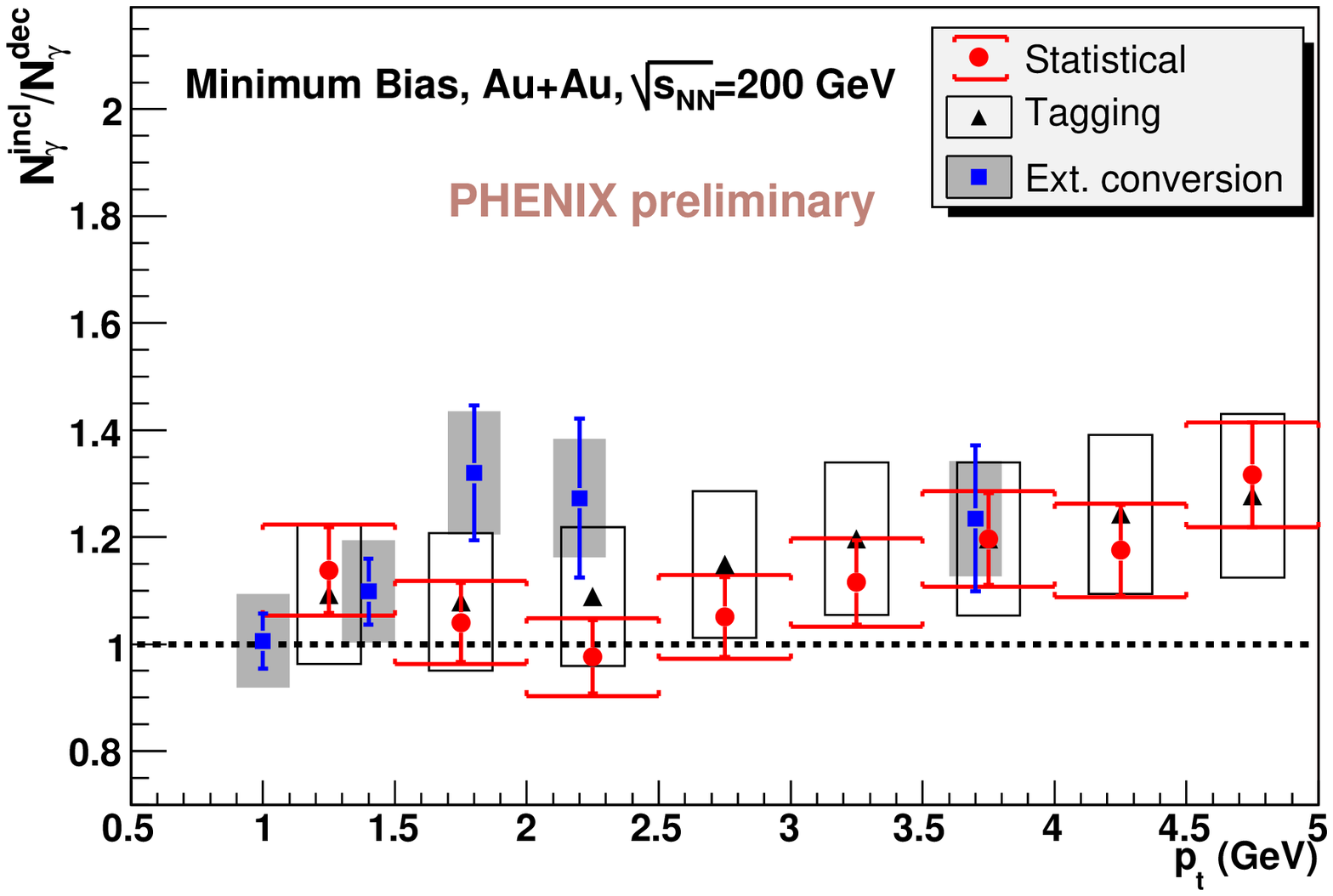}\hfill
  \includegraphics[width=75mm]{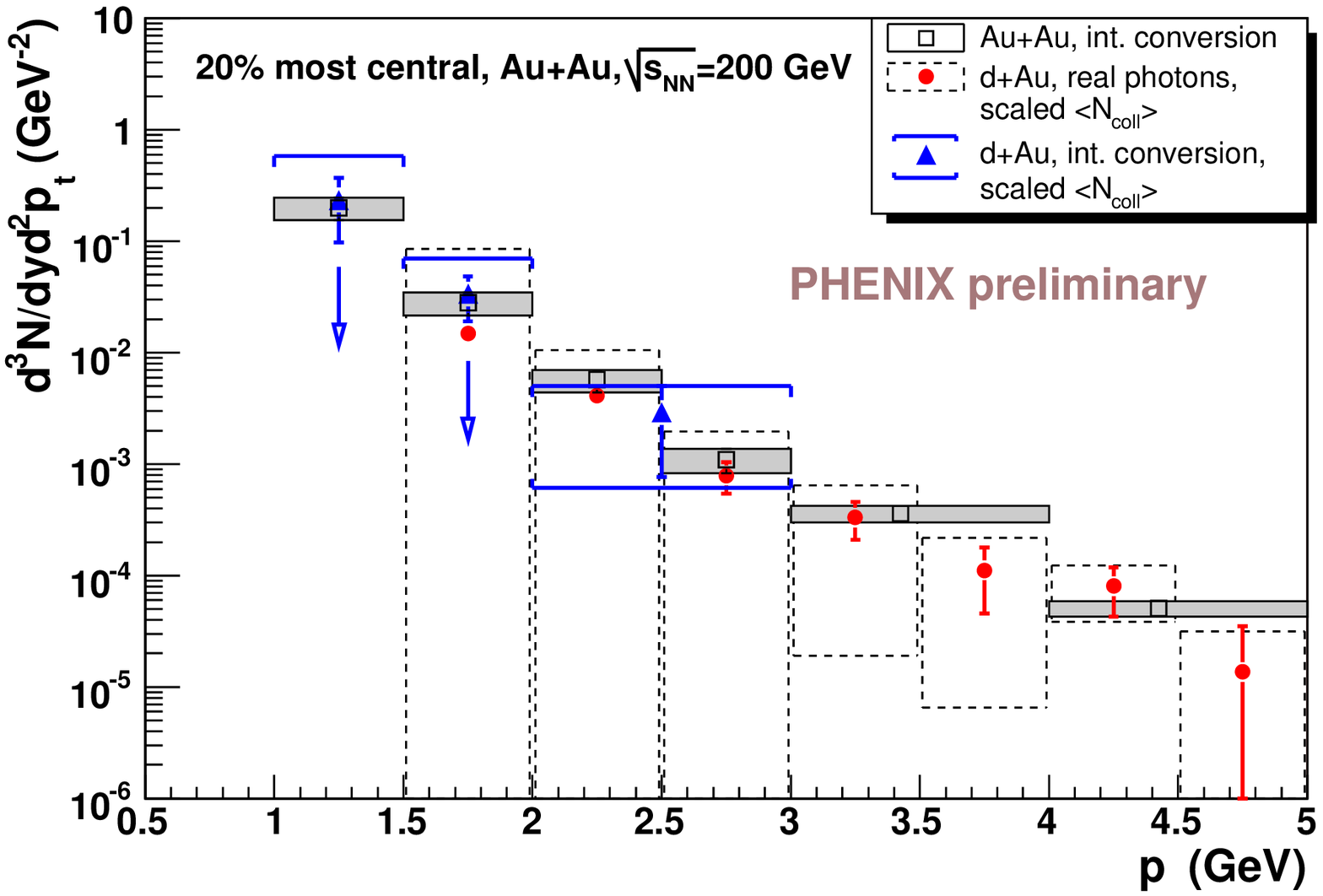}\\
  \caption{Left plot: Ratio of inclusive to decay photon yields measured in
  minimum bias Au+Au collisions at $\sqrt{s_{NN}}=200$ GeV by several methods.
  Right plot : The direct photon spectrum measured in the 20\% most central
  Au+Au collisions at $\sqrt{s_{NN}}=200$ GeV to the binary scaled direct
  photon yield in d+Au collisions at the same energy.}
  \label{fig:AuAu}
\end{center}
\end{figure}

The direct photon spectra measured in p+p and d+Au collisions at
$\sqrt{s_{NN}}=200$ GeV is compared with binary scaled NLO pQCD
predictions \cite{Vogelsang} in Figure~\ref{fig:ratio}. In both cases
good agreement with theory is found over the entire $p_T$ range
suggesting an absence of strong nuclear effects in d+Au collisions.
The ratio of inclusive photons over decay photons in Au+Au collisions
was extracted using three different methods, see
Figure~\ref{fig:AuAu}. All three methods produce consistent results
and comparable systematic errors while external conversion produces
the smallest errors at small $p_T$. Taking the direct photon yield
measured in d+Au collisions as a baseline we can estimate the prompt
hard photon contribution in Au+Au collisions and search for excess
direct photons radiated from the hot matter. In Figure~\ref{fig:AuAu}
the direct photon yield in the 20\% most central and in minimum bias
Au+Au collisions is compared to the scaled d+Au collision results.
Unfortunately, the systematic errors on the d+Au results are too
large to draw a meaningful conclusion about a possible excess
contribution due to thermal radiation.

\section{Conclusions}

We present the measured direct photon yields in p+p, d+Au, and Au+Au
at $\sqrt{s_{NN}}=200$ GeV. PHENIX has developed a variety of methods
to extract the direct photon yields and all provide consistent
results. Both in p+p and d+Au collisions we find agreement with
binary scaled NLO pQCD predictions within errors. The binary
collision scaled yield in d+Au collisions is in agreement with the
yield measured in Au+Au collisions. The large systematic errors do
not allow to draw conclusions on the presence of a direct photon
excess due to matter-related emission in Au+Au collisions.

\end{document}